# HIDDEN INFORMATION AND REGULARITIES OF INFORMATION DYNAMICS III

*Vladimir S. Lerner*
13603 Marina Pointe Drive, C-608, Marina Del Rey, CA 90292, USA, lernervs@gmail.com

**Abstract**

This presentation's Part 3 studies the evolutionary information processes and regularities of *evolution dynamics*, evaluated by an entropy functional (EF) of a random field (modeled by a diffusion information process) and an informational path functional (IPF) on trajectories of the related dynamic process (Lerner 2012).

The integral information measure on the process' trajectories accumulates and encodes inner connections and dependencies between the information states, and contains more information than a sum of Shannon's entropies, which measures and encodes each process's states separately.

Cutting off the process' measured information under *action of impulse controls* (Lerner 2012a), extracts and reveals *hidden information,* covering the states' correlations in a multi-dimensional random process, and implements the EF-IPF *minimax variation principle* (VP).

The approach models an information observer (Lerner 2012b)-as an extractor of such information, which is able to convert the collected information of the random process in the information dynamic process and organize it in the hierarchical information network (IN), Part2 (Lerner, 2012c).

The IN's highest level of the structural hierarchy, measured by a maximal quantity and quality of the accumulated cooperative information, evaluates the observer's *intelligence level*, associated with its ability to recognize and build such structure of a meaningful hidden information.

The considered evolution of optimal extraction, assembling, cooperation, and organization of this information in the IN, satisfying the VP, creates the phenomena of an evolving observer's intelligence.

The requirements of preserving the evolutionary hierarchy impose the restrictions that limit the observer's intelligence level in the IN. The cooperative information geometry, evolving under observations, limits the size and volumes of a particular observer.

## Part 3. The Evolutionary Regularities of Informational Dynamics

### *Introduction*

The studied evolution includes both progressive improvement of existing information systems with their extensive development and creation of more advanced systems, having enhanced evolutional properties, organized it in the hierarchical information network (IN), Part 2 (Lerner, 2012b), which concentrates all measured information. Information, as a logarithmic comparative measure of the compared states (events), in an *observed process* ought to measure its states' inner connections, integrating them through all process.

The integral information measure on the controlled Markov diffusion process' trajectories (Part 1) accumulates and encodes inner connections and dependencies between the information states, and contains more information than a sum of Shannon's entropies, which measures and encodes each process's states separately.

Cutting off the Markov diffusion process under *action of impulse controls* (Lerner 2012a) (associated with "killing its drift"), selects a minimal Markov path, where the states' inner connections concentrate a substantial amount of their hidden information, covering the states' correlations in a multi-dimensional random process. Applying the integral information measure during the cutoff evaluates this *minimal* information path, while the impulse control extracts a *maximum* of this information, thereby implementing *minimax variation principle* (VP). Thus, the approach not only reveals hidden information of the controlled process' inner connections, but also allows measuring its maximum *where* this information is concentrated and describing the process' *dynamics* which disclose its regularities. Hence, information of the observer's process actually is its *integrated hidden information*.

Such a hidden information could release a missing connections during, for example, creation of a human thought, speech, discussion, communication, and so on, being important for evaluation the process' *meaningful* information, related to these connections in acceptance, cognition, and perception of the information.

In the IN hierarchy of the extracted hidden information, each of its following higher level sequentially accumulates and binds more concentrated information that the previous one.

This evolving "depository" of maximal information is an observer's "intelligence center", which growths successively (through the VP implementation) during the interactive exchanges with environment.

The considered evolution of optimal extraction, assemble, cooperation, and organization of this information in the IN, satisfying the EF-IPF minimax VP, creates the phenomena of an evolving observer's intelligence.

However the requirements of preserving the evolutionary hierarchy impose the restrictions that limit the observer's intelligence level in the IN.

The VP establishes the following evolution information mechanisms of:

(1)- an optimal extraction of maximal available random information, collected by EF;

(2)-optimal information dynamics of the collected information, minimized by IPF.

It has shown that the VP single form of the mathematical law leads to specific evolutionary regularities, such as:

-creation of the order from stochastics through the evolutionary macrodynamics, described by a gradient of dynamic potential, evolutionary speed and the evolutionary conditions of a fitness and diversity;

-cooperative information dynamics, creating the group's units (doublets, triplets), enable maximize their minimal information for a mutual competing with others groups;

-the evolutionary hierarchy of the group's units in cooperative information network (IN), unified through growing their minimax information values and a potential of adaptation;

- the IN's evolving information geometry with the evolution of its node's space time-locations, which limit the size and volumes of a particular observer;

-the adaptive self-controls and a self-organization with a mechanism of producing and copying a genetic code for a cyclic reproduction, possessing an increased ability of growing the observer's intelligence.

This law and the regularities determine *unified functional informational mechanisms* of evolution dynamics.

For a real diffusion process with diffusing particles, the law is materialized by the particles' elementary interactions; and the law's evolution regularities will be implemented through real physical, chemical, biological structures, which for highly organized systems would create a cognition and intelligence.

In spite of numerous publications on evolution theory Dawkins 1976, Dyson 1999, Hedrick 2005, Joyce 1992, Kastler 1967, Kimura 1983, Michod 1999, Nicolis and Prigogine 1977, Wright 1968-69, 1977-78, others, the central questions concerning the existence of a general evolutionary law for Life, remains unanswered.

E. Schrödinger 1944, analyzing a physical aspect of Life, concludes that an organism supports itself through orders that it receives from its environment via the *maximization of a negative entropy*, as a general principle, connecting its dynamics and stochastics. Darwinian law's modern formula, Mayr 2004, stating that "evolution is a result of genetic variances through ordering by the elimination and selection", focuses on a competitive "*struggle* for Life" amongst organisms. G. Dover 2000 emphasizes on the genes' *cooperative* phenomena: "the central feature of evolution is one of tolerance and cooperation between interacting genes and between organisms and their environment… Genes are born to cooperate". Nowak 2006 proposes five mechanisms by which cooperation arises in organisms ranging from bacteria to human beings.

Are there any connections between all these concepts and formulas, covering them under a general law?

Most publications in mathematical biology, Murray 2002, Turchin 2003, do not provide a quantitative prognosis of the law phenomena and only support a law by experimental and/or simulation data.

For example, Schwammle and Brigatti 2003 describe the result of *simulation*, depending on priory *chosen* parameters, without a detailed insight of the *dynamic* mechanism of micro-macroevolution.

In such approaches, many essential evolutionary phenomena are missing.

A principal aspect of evolutionary law is an ability to *predict* the process development and phenomena based on the law mathematical forms. The existing mathematical formalism does not satisfy these requirements; the evolution equations are not bound by a general principle related to a unique law. The modern evolution theory is dominated by diverse assumptions, concepts, methods, and hypotheses. The unique formalism of evolutionary law, as well as general systemic regularities of evolution, expressed in information form, are still unknown.

We generalize different forms of evolution process considering a *flow of information* as an *information process independently of its (physical ,biological, economical, other) origin;* the flow is *freely distributing in a bounded space–time environment, which is randomly affecting the flow*.

Such approach focuses on the *process' information dynamics and their creation from stochastic dynamics.*

The questions are: What are the regularities of this process, expressed in an information form? Can these regularities be revealed and extracted through the process' observations? What are the most informative space-time observations of this flow (providing a maximum of extracted information to reveal the seeking regularities)? Which principle could be used to solve these problems? What are limitations on evolution dynamics?

Because of the universal nature of information, such an information approach is *applicable* to some physical processes, represented by their information models (Lerner 1999,2008,2011).

This Part's goal is to describe *general systemic regularities* of evolutionary process, based on *information dynamic* approach with a single variation principle (VP) as a mathematical law, and *apply* them to an information observer (Lerner 2012b). The evolution dynamics proceeds under the VP generated controls, which provide directional consolidation of the extremal segments' dynamics in the cooperative structures at each "window" between the segments. At the window, external stochastics affect the dynamics, leading to the process renovation.

We show that instead of a "punctuated equilibrium" (Gould and Eldredge 1993), evolution depends on a punctuated *nonequilibrium* rising at the interactive windows, where the microlevel's stochastics contribute to the macrolevel's dynamics, generating a dynamic potential of evolution.

The applied variation principle leads to a dynamic model of *open system* with an *irreversible* macroprocess, which originates from the random microprocess through its entropy integral (EF).

The known *variation* problems produce the equation of a *close* system with a *reversible* processes.

### 3.1. The impact of the initial stochastics on macrodynamics

The IMD transforms of the initial random motion into the dynamic motion, proceeding on a sequence of extremal segments, which are selected from the related segments of the random process with a maximal probability; while the random windows between the extremal segments provides a current information needed for the dynamic motion on each following segment. A sum of the renovated extremal segments forms the trajectory of irreversible macroprocess, which approximates the initial random process with a maximal probability (Part 1).

Following the VP, the extremal segments are sequentially assembled in triplets, while each next triple includes the previous one, maintaining a concentration of information from all preceding triplets.

The triplet's cooperative dynamics with the related windows build a *cooperative hierarchical information network* (IN) (Fig.2.4), where the initial random motion develops the triplet's cooperative dynamics, concentrating in the IN nodes. Each triplet's node accumulates both information dynamics from the node's three segments and information from the random windows between each of them. Such information, inherited by the current triplet's dynamics, evolves in the following triplet's dynamics through the *o*-window between the triplets.

The window's information renovates the inherited information by analogy with mutation. The triplet's cooperative organization in the IN is *an inner the ability* of the information cooperative dynamics (Parts 1,2).

We study the evolution regularities of such a cooperative, focusing on an impact of the initial stochastics on the produced dynamics.

Since the dynamic motion from the IN's *m* nodes is concentrated in the IN's final node, its dynamic operator at the ending moment $t_m^{k=3m}$ of the last segment's dynamics: $\alpha_m(t_m^{k=3m})$ (defined by this eigenvalue of the model operator), is connected with the dynamic operator of a first IN's node, at the ending moment $t_{m=1}^{k=3}$ of this segment's dynamics: $\alpha_1(t_{m=1}^{k=3})$ by relation

$$\alpha_m(t_m^{k=3m}) = \alpha_1(t_{m=1}^{k=3})\gamma_2^{-m}, \tag{1.1}$$

where $\gamma_2$ is the IN parameter which each triplet's ending eigenvalue (Part 2).

Because the above dynamic operators by the end of each segment: for the IN first $\alpha_1(t_{m=1}^{k=3})$ and its last *m*-segment $\alpha_m(t_m^{k=3m})$ accordingly, and the related dynamic operators at the segment's beginning: $\alpha_1(t_{1o}^{3k=1})$ and $\alpha_m(t_{mo}^{k=3m})$, satisfy the same invariant equalities:

$\alpha_m(t_{mo}^{k=3m}) = \mathbf{a}(\gamma)\,\alpha_m(t_m^{k=3m})$, and $\alpha_1(t_{m=1}^{k=3}) = \mathbf{a}(\gamma)\alpha_1(t_{1o}^{k=3})$,

where $\mathbf{a}(\gamma)$ is the model invariant, $\gamma$ is the model's basic parameter Part1,

we come to the same ratio (as (1.1)) for the segment's starting eigenvalues:
$$\alpha_m(t_{mo}^{k=3m})/\alpha_1(t_{1o}^{3k=1}) = \gamma_2^{-m}. \tag{1.1a}$$
Let us find the relative changes of correlation along the dynamic process, evaluated by the ratios (Part1):
$$\dot{r}_1^* = dr_1(t_1)/r_1(t_1)dt\big|_{t=t_{1o}} = 2\alpha_1(t_{1o}), \dot{r}_m^* = dr_m(t_m)/r_m(t_m)dt\big|_{t=t_{mo}} = 2\alpha_m(t_{mo}), \tag{1.2}$$
where $\dot{r}_1^*$ evaluates a relative speed of correlation at beginning of the IN first segment, which determines the eigenvalue $\alpha_1(t_{1o}^{k=1})$ at the beginning of time interval $t_{1o}^{k=1}$ of this segment. This eigenvalue arises through collecting the correlations of a random process during *first o*-window; while $\dot{r}_m^*$ evaluates a relative speed of correlation at beginning of the IN last *m*-segment that determines the eigenvalue $\alpha_m(t_{mo}^{k=3m})$ at the beginning of the time interval $t_{mo}^{k=3m}$ of this segment. This eigenvalue arises through collecting the correlation from a random process during a *last o*-window. Therefore, the ratio
$$\dot{r}_m^*/\dot{r}_1^* = \alpha_m(t_{mo})/\alpha_1(t_{1o}) = \gamma_2^{-m} \tag{1.3}$$
evaluates a relative decline of quantity of information, collected from the random process by its end, through all $3m$ windows for IN's *m* nodes.

This ratio estimates a relative impact of the random process on its information dynamics.

It's seen that this information impact essentially weakens by the process' end.

This means that the process' ending dynamic operator becomes less dependable on the initial process' randomness, being determined mostly by the macrodynamics, accumulated sequentially in the IN's ending node.

Otherwise, the cooperative evolution process is determined by the VP *dynamic* law rather than by the initial *random* process, whose impact during the evolution *diminishes*.

The randomness is reduced under emergence of the model's three dynamic cooperative mechanisms (Parts 1,2):

1. Internal informational macrodynamics at imposing the constraint;

2. Quantum information dynamics on the edges of the windows where random information accesses;

3. The IN hierarchical cooperative dynamics, generating the information forces between the IN's units.

The evolutionary dynamics of each following IN node enfolds and minimizes total random changes from all preceding nodes. While a single information *unit* is seeking for a cooperation.

From these, it follows that *cooperative dynamics, enfolding less IN's nodes, undergo more external influences with a potential impact on its evolution dynamics than cooperative system that enfold more IN's nodes.*

### 3.2. The reversible and irreversible time intervals

The evolutionary model possesses two scales of time: a reversible time equals to a sum of the time intervals on the *extremals* segments: $\sum_{i=1}^{n} t_k^i = T_e^r$, and the irreversible time, counted by a sum of the irreversible time elementary intervals *between* the segments $\delta t_i^{ir}$:

at $\delta t_i^{ir} = (\Delta S_i^\delta / \mathbf{a}_{oi}^2(\gamma) - 1)t_i^r,$ \hfill (2.1)

where $t_i^r$ is a regular time interval determines by the segment's dynamics during its each $t_k^i$ (Parts1,2). (Proof of (2.1) is in (Lerner 2010).

Here $\Delta S_i^\delta$ is an elementary information contribution between the segments, generated by a random process,

$$\mathbf{a}_{oi} = \alpha_{io} t_i^r \qquad (2.1a)$$

is the segment's dynamic invariant and $\mathbf{a}_{oi}^2$ is the invariant's evaluation of the model's impulse control. A sum

$$T_e^r + T_\tau^{ir} = T_m^l \qquad (2.2)$$

defines the model lifetime.

If each $\Delta S_i^\delta$ between the segments is compensated by the impulse control action with $\Delta S_i^\delta = \mathbf{a}_o^2$, then the irreversible time does not exist, means that the macroprocess' dynamics cover the random contributions, which are measured by $\mathbf{a}_{oi}^2$. At any $\Delta S_i^\delta > \mathbf{a}_o^2$, the stochastics affect dynamics bringing the macroprocess' irreversibility.

Both reversible and irreversible time intervals depend on the quantity and quality of information, accumulated within these discrete, which is measured by the considered information invariants (Part2).

Since $\delta t_i^{ir}$-locality involves quantum phenomena (Part 1), the connection of these times in such a locality reveals the model's *time's quantum informational nature*. Both of these times evolve with the evolution of the invariants and the system information dynamics (classical and quantum).

Applying invariant relations (2.1a) to (1.3) we get the ratio of reversible time interval at the beginning $t_{1o}$ and the end $t_{mo} = T_{m=n/2}$ of the optimal process:

$$t_{mo} / t_{1o} = \gamma_2^m . \qquad (2.3)$$

The ratio of corresponding irreversible time intervals (2.1) is not changing during the considered consolidation process, while with growing $t_k^i$ (at the macrolevel), the relative influence of quantum time locality $o_t(\tau_k^{i1})$ stays the same, acting proportional to $t_k^i = t_i^r$. Therefore, the process' total irreversible time:

$$T^{ir} = \delta t_{1o}^{ir} \sum_{i=1}^{m=n/2} \varepsilon_i = \delta t_{1o}^{ir} \varepsilon_i n / 2 \qquad (2.4)$$

increases proportional to the IN triplet's number $m = n/2$, while a total process' reversible time increases much faster according to $T^r = t_{1o} \gamma_2^m$. Their ratio: $T^r / T^{ir} = 2\gamma_2^m \varepsilon_i^{-2} m^{-1}$ at $t_{1o} / \delta t_{io}^{ir} = \varepsilon_i^{-1}$, $\varepsilon^2 \cong 0.0675$ (Part 2), also increases: $T^r / T^{ir} \cong \gamma_2^m 30 / m$, taking the values $T^r / T^{ir} \cong 1.717$ at $m = 4$ and $T^r / T^{ir} \cong 196.482$ at $m = 8$.

This means, the optimal process evolves according to the *regular* time course with growing dimensions of its cooperative structures.

### 3.3. The main limitations on the evolution macrodynamics
#### 3.3.1. Limitations on the invariants and their evolution

Since each elementary information contribution between the segments, generated by a random process, is measured by invariant $\mathbf{a}_o^2(\gamma)$, *maximum* of this information satisfying the VP, corresponds to

$$max \, \mathbf{a}_o^2(\gamma) \rightarrow \Delta S_i^\delta = \mathbf{a}_o^2, \qquad (3.1)$$

which is reached at $\gamma \rightarrow 0$.

Both real invariant $\mathbf{a}_{oi} = \alpha_{io} t_i^r$ and imaginary invariant $\mathbf{b}_{oi} = \beta_{io} t_i^r$ (defined through of the model's complex eigenvalues $\lambda_{io} = \alpha_{io} \pm j\beta_{io}$ of its dynamic operator) depend on $\gamma$ according to Eq:

$$2(\sin(\gamma \mathbf{a}_o) + \gamma \cos(\gamma \mathbf{a}_o)) - \gamma \exp(\mathbf{a}_o) = 0 \qquad (3.1a)$$
$$2\cos(\gamma \mathbf{b}_o) - \gamma \sin \gamma(\mathbf{b}_o) - \exp(\mathbf{b}_o) = 0 \qquad (3.1b)$$

Invariants $\mathbf{a}(\gamma)$, which measures information contributions from a regular control, is connected with $\mathbf{a}_o(\gamma)$ by Eq:

$$\mathbf{a}(\gamma) = \exp(-\mathbf{a}_o(\gamma))(1-\gamma^2)^{1/2} \times [4 - 4\exp(-\mathbf{a}_o(\gamma))\cos(\gamma \mathbf{a}_o(\gamma)) + \exp(-\mathbf{a}_o(\gamma))]^{-1/2}, \ \mathbf{a}_o(\gamma) > 0, \ (3.1c)$$

At $\gamma \to 0$, Eqs (3.1a-3.1c) have the following solutions for these invariants:
$\mathbf{a}_o(\gamma \to 0) \cong 0.7652, \mathbf{a}(\gamma \to 0) \cong 0.2328, \mathbf{b}_o(\gamma \to 0) \cong 0.7$.

We evaluate also the nearest solutions for these invariants: $\mathbf{a}_o(\gamma \to 1) \cong 0.3, \mathbf{a}(\gamma \to 1) \cong 0.3, \mathbf{b}_o(\gamma \to 1) \cong 0.3$.

The condition of *cooperation* (for the triplet's ratios $\gamma_1^\alpha(\gamma), \gamma_2^\alpha(\gamma)$ in Part 2) limits $\gamma$ admissible for the triplet's cooperative dynamics by $\gamma \to (0.0-1.0)$. The restricted ratio of eigenvalues $\gamma_1^\alpha = \gamma_2^\alpha = 1$ leads to repeating the model's eigenvalues (that contradicts the initial assumptions of the information process' nonredundancy). This limits the values for all invariants and constrains the model's dynamics.

Indeed. In a chain of connected information events (states), the appearance of an event, carrying $\gamma \to 1$, leads to $\mathbf{a}_o(\gamma=1)=0$ when both information contributions from the regular control $\mathbf{a}(\gamma=1)=0$ and the impulse control $\mathbf{a}_o^2(\gamma=1)=0$ turn to zero. This indicates impossibility of delivering external information to the dynamic model, dissolving the dynamic constraint (Part 1) that disintegrates of the dynamics.

At a locality of $\gamma=1$, both the event's information $\mathbf{a}_o$ and the related time undergo a jump, which could be the indicators of approaching $\gamma=1$. For example, when $\gamma$ approaches 1, $\mathbf{a}_o(\gamma)$ is changing from $\mathbf{a}_o(\gamma=1-o)=0.56867$ to $\mathbf{a}_o(\gamma=1)=0$, and the corresponding time's ratio of the following and preceding intervals reaches $\tau_{i+1}/\tau_i = 1.8254$ with the eigenvalues' ratio $\alpha_{io}/\alpha_{it} \cong -1.9956964$.

This jump is a *dynamic indicator* of breaking up the dynamic constraint, which leads to cutting off the model's dynamics from the initial random process with a possibility of getting more uncertainty and rising chaotic dynamics. Thus, the appearance of an event, carrying $\gamma \to 1$, leads to a *chaos and decoupling* of the chain, whereas the moment of this event's occurrence could be *predicted* by measuring a current event's information $\mathbf{a}_o$ and using it to compute $\gamma$. The evolution of the invariants at $\gamma \to (0 \rightleftharpoons 1)$ is shown on Fig. 3.1.

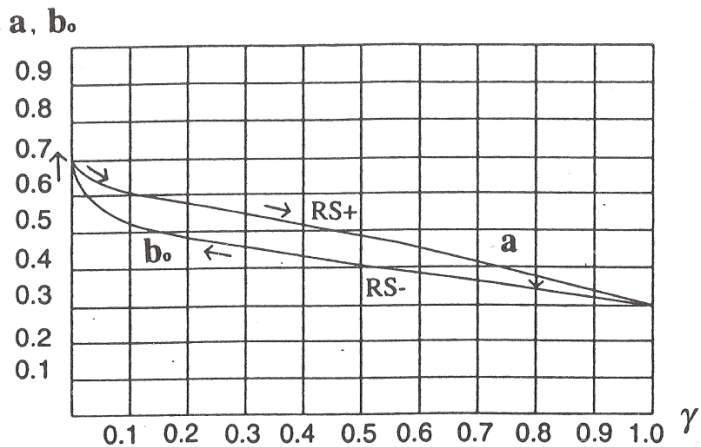

Figure 3.1. The evolution of the model invariants with a loop between information subspaces for the invariants' real RS+ and imaginary RS− values.

The macromodel evolves toward a minimal $\gamma \to 0$, while the evolution increases the number $m$ of consolidated triplets' subsystems. With increasing $\gamma \to 1$, $m$ decreases, whereas the actual region of admissible $\gamma$ (that excludes above singularities) is limited by $\gamma \to (0.0-0.8)$.

The consolidation process, preserving invariant $\mathbf{a}_o$, develops along the line of switching controls, corresponding $\mathbf{a}(\gamma) = |\alpha_i^t t_i| = inv$, which evaluates a region with geometrical locality (RS+) of a real entropy.

The macrosystem can evolve along this line at a constant total entropy $S = S_e + S_i$ if its internal entropy ($S_i > 0$) is compensated by an external negentropy ($S_e < 0$), delivered from the environment. Until the entropy production is positive ($d\Delta S_i / dt > 0$), the time course is also positive, and the entropy is a real $S^a$.

Another character of the macrodynamics takes place along a line $j\beta_{n+k}^i(t_{n+k})t_{ik} = \beta_{n+k}^i(t_{ik})t_i^-$, where $t_i^- = jt_{ik}$ is an imaginary time in a region with geometrical locality (RS−) of an imaginary entropy $S^b$, evaluated by $S^b = |\beta_{n+k}^i(t_{n+k})t_{ik}|$. The second law along this line, as well as any real physical law, is not satisfied.

Considering the start of the imaginary movement (by the end of region RS+) at some moment $t_{n+k-1}$, when $\gamma_n = |\beta_{on}^i / \alpha_{on}^i| \cong 1$ and $\beta_{on}^i = \beta_{n+k-1}^i$, $|\alpha_{on}^i| \cong |\beta_{n+k-1}^i|$, we get the ratio $S^b / S^a = |\beta_{n+k}^i(t_{n+k})t_{ik}| / |\alpha_o^i t_i|$, which, assuming $t_{ik} \cong t_i$, acquires the form

$$S^b / S^a = |\alpha_{on}^i| / |\alpha_o^i| = h_\alpha. \tag{3.2}$$

This ratio we can evaluate by a maximal ratio of the invariants:

$$\mathbf{a}_o(\gamma \to 1) / \mathbf{a}_o(\gamma \to 0) \cong 0.4, \tag{3.2a}$$

estimating $\mathbf{a}_o(\gamma \to 1) \approx \mathbf{a}_o(\gamma \cong 0.8)$ and $\mathbf{a}_o(\gamma \to 1) \cong \alpha_{on}^i t_i$.

Considering a minimal achievable ratio for invariants $h_\alpha$: $\min h_\alpha = h_\alpha^o$, we evaluate a current invariant's ratio $h_\alpha$ (3.2) by the value $h_\alpha = h_\alpha^o + \Delta h_\alpha$, which we use to estimate the entropy ratio (3.2):

$$S^b / S^a = h_\alpha^o(1 + \Delta h^*), \Delta h^* = \Delta h_\alpha / h_\alpha^o. \tag{3.3}$$

The controlled macroprocess, after a complete cooperation, is transformed to one dimensional consolidated process $x_n(t) = x_n(t_{n-1})(2 - \exp(\alpha_{n-1}^t t))$, which at $t_{n-1} = \dfrac{\ln 2}{\alpha_{n-1}^t}$ is approaching the final state $x_n(t_n) = x_n(T) = 0$ with an infinite phase speed

$$\frac{\dot{x}_n}{x_n}(t_n) = \alpha_n^t = -\alpha_{n-1}^t \exp(\alpha_{n-1}^t t_n)(2 - \exp(\alpha_{n-1}^t t_n))^{-1} \to \infty. \tag{3.4}$$

However, the model cannot reach the zero final state $x_n(t_n) = 0$ with $\dot{x}_n(t_n) = 0$, since, at approaching it, a periodical process arises, as a result of alternating the movements with the opposite values of each *two* relative phase speeds $\dfrac{\dot{x}(t_{n+1})}{x_n(t_{n+1})}, \dfrac{\dot{x}(t_{n+2})}{x_n(t_{n+2})}$ and under the related control switches. The process is limited by admissible error $\varepsilon^* = \dfrac{\Delta x_n}{x_n}$ and/or the related time deviations of switching control $\varepsilon_m^* \leq \Delta t_i / t_i = \alpha_{io}^*$.

This result *limits a possibility of a total completion the model's cooperation*, which should be restricted by a minimum of two non-cooperated IN's ending eigenvalues.

### 3.3.2. Model's uncertainty and its limitations

Each *o*-window between the extremal segments is a source of uncertainty generated by the random microlevel, while its local uncertainty exists at each locality of a triplet's node, where two extremal segments join a third segment.

By closing the $o$-window, the optimal controls overcome uncertainty, allowing connecting the segments and initiating a sequential time-space movement from each previous to a following IN node.

Along such a movement, a zone of uncertainty UR, surrounding each IN node, is formed (Fig.2.5), where its double and then triple cooperation takes place.

A local uncertainty (between any two segments) $UR_i$ is evaluated by a relative error

$$\varepsilon_i(\gamma) = \delta t_i / t_i(\gamma_i^\alpha(\gamma)) \to \delta t_i^{ir} / t_i^r , \qquad (3.5)$$

which also estimates the relative irreversible time interval (2.1).

The error decreases through binding the eigenvalues at cooperation. For example, by connecting first two eigenvalues of a triplet, we get $\gamma_{1,2}^\alpha = 2.215$ (at $\gamma = 0.5$) and $\varepsilon_1 \leq 0.078$. By adjoining them with a third eigenvalue of the ranged spectrum, we get $\gamma_{2,3}^\alpha = 1.756$ (at $\gamma = 0.5$), and $\varepsilon_2 \leq 0.07$, at $\gamma_{1,3}^\alpha = \gamma_{1,2}^\alpha \gamma_{2,3}^\alpha = \gamma_2^\alpha = 3.89, \gamma_{1,2}^\alpha = \gamma_1^\alpha$.

Since each cooperation takes place in UR, the number of cooperations $m_c$ is determined by number of the IN's $UR(\varepsilon_i(\gamma))$, while the existence of each $\varepsilon_i(\gamma)$ depends on $\gamma$, changing within the limited interval $\gamma \to (0.0 - 0.8)$.

Let us find a minimal interval of changing $\gamma$, defined by increment $\gamma^* - \gamma_o = \Delta \gamma_o = \gamma^* - (\gamma_o = 0) = \gamma^*$.

This increment will be determined by creation of a single cooperation, which changes the model's dimension $m_o(\gamma_o)$. While an elementary negentropy ($\mathbf{a}_o^2(\gamma) + \mathbf{a}(\gamma)$) is delivered on an interactive window for each extremal segment (Part 2), the segment consumes an elementary quantity of information $\mathbf{a}_o(\gamma)$.

The difference $\mathbf{a}_o^2(\gamma) + \mathbf{a}(\gamma) - \mathbf{a}_o(\gamma) = \delta \mathbf{a}(\gamma)$, as a surplus, can be used for forming particular information cooperation, and the ratio

$$m_c(\gamma) = (\mathbf{a}_o^2(\gamma) + \mathbf{a}(\gamma))/(\mathbf{a}_o^2(\gamma) + \mathbf{a}(\gamma) - \mathbf{a}_o(\gamma)) \qquad (3.6)$$

characterizes the number $m_c$ of such potential cooperations.

At $\gamma = 0$, we have $m_o = m(\gamma = 0) = (\mathbf{a}_o^2(\gamma = 0) + \mathbf{a}(\gamma = 0))/(\mathbf{a}_o^2(\gamma = 0) + \mathbf{a}(\gamma = 0) - \mathbf{a}_o(\gamma = 0))$, (3.6a)

and at $\gamma = \gamma^*$, we get

$$m(\gamma^*) = m_1 = (\mathbf{a}_o^2(\gamma^*) + \mathbf{a}(\gamma^*))/(\mathbf{a}_o^2(\gamma^*) + \mathbf{a}(\gamma^*) - \mathbf{a}_o(\gamma^*)). \qquad (3.7)$$

Since an increase of $\gamma$ from $\gamma_o = 0$ has a tendency of diminishing cooperation, we determine such a minimal $\gamma^*$ that needs just for a single cooperation.

This cooperation will decrease the initial model's dimension $m_o$ on 1, requiring

$$m_o - 1 = m(\gamma^*) . \qquad (3.7a)$$

Applying the formula (3.1c), connected both invariants, we have at $\gamma_o = 0$:

$$\mathbf{a}(\gamma) = \exp(-\mathbf{a}_o(\gamma_o))[4 - 4\exp(-\mathbf{a}_o(\gamma_o)) + \exp(-2\mathbf{a}_o(\gamma_o))]^{-1/2}. \qquad (3.7b)$$

By substituting $\mathbf{a}_o(\gamma = 0) = 0.76805$ we get the corresponding $\mathbf{a}(\gamma = 0) = 0.231960953$ and $m_o = 15.2729035$.

Using the same formulas, connected $\mathbf{a}_o(\gamma^*)$ and $\mathbf{a}(\gamma^*)$, which satisfy both (3.6a), (3.7) and also (3.7a), we come to equation $15.2729035 - 1 = m_1$, $m_1 = f(\mathbf{a}_o(\gamma^*))$, from which we get $\mathbf{a}_o(\gamma^*) = 0.762443796$ and

$\mathbf{a}(\gamma^*) = 0.238566887$. These values impose *upper limit* on the invariants at $\gamma^* > \gamma_o = 0$.

Above relations bring $m_1 = 14.2729035$ and evaluate $\gamma^* = 0.007148$.

This allows us also evaluate the UR information *border* by the relative increment of the extremal segment's entropy (3.2, 3.3), (concentrated in UR) through the model's invariants:

$$h_\alpha^o = (\mathbf{a}_o(\gamma=0) - \mathbf{a}_o(\gamma^*))/\mathbf{a}_o(\gamma=0) \qquad (3.8)$$

at changing $\gamma$ from 0 to $\gamma = \gamma^*$, where $\gamma^*$ corresponds to the segment's location at the UR border.

Here $h_\alpha^o$ is an *invariant*, determined by the increment of maximal invariant $\mathbf{a}_o(\gamma=0)$ regarding the information invariant $\mathbf{a}(\gamma^*)$, which measures a *minimal elementary* uncertainty *separating the model's dimensions.*

This $h_\alpha^o$ estimates a minimal information, needed for creation of single element (in a double cooperation), which then can start producing the IN information *structure*. We call $h_\alpha^o$ the model's *structural invariant*.

From above invariants we get

$$h_\alpha^o = 0.00729927 \cong 1/137, \qquad (3.8a)$$

which coincides with the Fine Structural constant in Physics (Krane 1983):

$$\alpha^o = 2\pi \frac{e^2}{4\pi\varepsilon^o hc}, \qquad (3.9)$$

where $e$ is the constant of electron's charge magnitude, $\varepsilon^o$ is the permittivity of free space constant, $c$ is the speed of light, $h$ is the Plank constant. Both $\alpha^o$ and $h_\alpha^o$ are dimensionless.

The model's maximal physical information frequency $\lambda^o$ has a meaning of maximal frequency for energy spectrum $v_{max} \cong 2.82 k\Theta/h$ (in unit of $[v_{max}] = \sec^{-1}$), where $h$ is Plank constant and $\Theta$ is absolute temperature. The related information frequency $\lambda^o$ (in $[\lambda^o] = Nats/\sec$) is equal to

$$\lambda^o = v_{max}/Nats = 2.82/hNats = \hat{h}^{-1}, \qquad (3.9a)$$

where $\hat{h}$ is an information equivalent of Plank constant. At a room temperature, we get

$$\hat{h} \cong 0.5643 \bullet 10^{-15}[\sec/Nats]. \qquad (3.9b)$$

The model's minimal uncertainty $h_s$ satisfies the relations

$$(\mathbf{a}_o(\gamma=0) - \mathbf{a}_o(\gamma^*)) \to h_s \cong 5.6 \bullet 10^{-3} Nats, \qquad (3.10)$$

and $h_s = h_\alpha^o \mathbf{a}_o(\gamma=0)$ (3.10a) that follows from (3.8, 3.8a).

Ratio $\hat{h}/h_s \cong 10^{-18}[\sec]$ evaluates *minimal time interval* of an information wave corresponding $\hat{h}$.

All four invariant's measures: $\alpha^o, h$ and $h_\alpha^o, h_s$ are connected by equalities $h_\alpha^o \cong \alpha^o$, (3.9), (3.9a), (310a), where $h_s$ estimates a *minimal elementary* uncertainty *(entropy)* needed for creation of a single information element, $h_\alpha^o$ *evaluates the same entropy relatively to the model minimal invariant, while* $h$ *is an energy's measure of* $h_s$, and $\alpha^o$ is a physical equivalent of $h_\alpha^o$ in a structure of elementary particles. The above equalities lead to connection of the IMD model to Quantum Mechanics QM (considered in Part 1 and Lerner 2012a).

The invariant quantity of information $\mathbf{a}_o(\gamma)$ has a distinctive information value depending on its location at the IN's hierarchy. This invariant also estimates the minimal achievable ratio $S^b/S^a \to h_\alpha^o$ in (3.2) at $h^* \to 0$, and it is limited by the minimal $\mathbf{a}_o(\gamma^*)$ in (3.7). The minimal $\Delta\gamma^* = \gamma^* - \gamma_o = \gamma^*, \gamma_o = 0$ that restricts reaching $\gamma = 0$, determines the quantities of information for invariants $\mathbf{a}_o(\gamma^*)$ and $\mathbf{a}(\gamma^*)$, which can generate an elementary negentropy $s_h(\gamma^*) = \mathbf{a}_o^2(\gamma^*) + \mathbf{a}(\gamma^*) = 0.8119887424$, while a maximal potential negentropy is $s_h(\gamma \to 0) = \mathbf{a}_o^2(\gamma \to 0) + \mathbf{a}(\gamma \to 0) = 0.8218$ at $\mathbf{a}_o(\gamma^*) = 0.762443796$.

Potential dimension $m_o$=15.2729035 is not achievable, because of existence of irremovable uncertainty, whose hidden information might produce just one cooperative element (doublet, or triplet).

Therefore, an real transfer from $\gamma_o = 0$ to $\gamma^* = 0.007148$ is able to create only a single cooperative element. Since the initial real dimension $m_{or}(\gamma_o = 0) \to 0$, by following (3.7a), we get $|m_{r1}| = 1$, while $m_1 = f(\mathbf{a}_o(\gamma^*))$ is used to find the invariants and $\gamma^*$. A single cooperative element, which is able to produce a maximum of "free information" $s_h(\gamma^*)$, could use its part $s_c(\gamma^*) = \mathbf{a}_o^2(\gamma^*) = 0.58132$ on cooperation with other elements, while it might apply another part $\mathbf{a}(\gamma^*) = 0.23$, as a control, to start a dynamic process on a joint element.

With growing $\gamma$ between $\gamma > \gamma^*$ and $\gamma \leq 0.8$, both free information $s_h$ and its part $s_c$, spent on cooperation, decline. Specifically we get

$s_h(\gamma = 0.3) = 0.79273, s_c(\gamma = 0.3) = 0.553, s_h(\gamma = 0.5) = 0.75, s_c(\gamma = 0.5) = 0.498436,$
$s_h(\gamma = 0.6) = 0.7269, s_c(\gamma = 0.6) = 0.4673, s_h(\gamma = 0.7) = 0.7033, s_c(\gamma = 0.7) = 0.435$
$s_h(\gamma = 0.8) = 0.6806, s_c(\gamma = 0.8) = 0.404.$

To achieve information balance, satisfying the VP, each single unit searches some partners for consumption-needed information.

A double cooperation conceals information $s_c(\gamma) = \mathbf{a}_o^2(\gamma)$, while a triple cooperation conceals information $s_{cm}(\gamma) = 2\mathbf{a}_o^2(\gamma)$ and it could produce less free information depending on $\gamma$.

With more triplets, cooperating in an IN, the cooperative information grows, while their free information is spent on joining each following triplet.

In a three dimensional space, above $\mathbf{a}_o(\gamma^*)$ provides information necessary to form a triplet $m_k^3 = 3$, which carries 4 bits of a genetic DSS code (Part 2). The minimal number $m_1$ (at this threshold) (3.7a) brings three-dimensional $m_1^3 \cong 42.8187105 \cong 43$ with a potential genetic code carrying 172 bits of a non-redundant information.

A total number of model's dimensions $m_c = n/2$ can produce 172 bits with code density $(\gamma_m^\alpha)^{m_c}$ (Part 2).

This means that *a non-removable uncertainly enfolds a potential DSS information code.*

Its minimal relative invariant $h_\alpha^o$ evaluates an elementary *increment* of the model's dimensions in (3.8), while a total quantity of hidden invariant information $s_h$ is able to produce an elementary triple code, enclosed into the hyperbolic cellular structure Fig.2.6.

A non-removable uncertainty is inherent part of any interaction (cooperation); it *evaluates non-separable connections of the states (events) in an information process as its elementary hidden information.*

The above consideration extends the VP application, allowing us to *conclude* that

(1)-*the model's invariants not only characterize the region between the model's real and imaginary eigenvalues, but also limit the UR;*

(2)-*the model's irreversible time emerges as an information measure of uncertainty for both each local o-window and the whole UR;*

(3)- *the UR can be measured in a discrete unit of uncertainty, expressed in the bits of the information code, as well as by the cell's number and sizes in the information geometry.*

The last result is connected to the notion of information as a measure of uncertainty, allowing us to introduce its geometrical representation and new meaning via the irreversible time.

Both the reversible time intervals $t_i = \mathbf{a}_o(\gamma)/\alpha_{io}^t$ and its equivalent of the scalar space intervals $l_i = \mathbf{a}_{mo}/\alpha_{io}^l$ are determined by the model's eigenvalue spectrum.

The entire UR's geometry identifies the IN's geometrical border where the macromodel is open for the external interactions. The conic structures, generating the UR cells, form an information geometrical background for the IN.

At the given cells' logic, the sequence of cells' blocks forming the IN nodes, can be built.

The above results impose *essential restrictions* on both a *general evolution progress* and *each step of the cooperative formation* that involves building not only an observer but also their cooperating collectives.

The restrictions include the limitations on:

- creation of these formations and accepting them in an existing cooperative;

-maximal units' number in a cooperative;

-the ability of a unit (or units) to cooperate with a particular collective, and so on.

Any cooperating units should benefit in cooperation through getting such a minimax information, which would be needed to overcome a threshold for a next cooperation. With growing the cooperative units, such threshold holds more condense information, enclosed in a high dimensional IN structure.

That is why forming the highly organized collectives requires the *increase of intelligence* that enables evolutionary grows of such complex cooperative systems.

It also follows that sustainable evolution changes, enable support an existing cooperative structure, are very limited, and the *limitations increase* with growing cooperative organization. The evolutionary changes on the IN's intelligence level could be possible through *particular variations of its DSS* code (Part 2).

The limitations increase the *information differences* between the nearest IN levels' subsystems, *restricting* formation such of them, which unable support the level's cooperative dynamics.

### 3.4. The potentials of the evolutionary dynamics

The information dynamics evolve within the above limitations.

The information speed of evolution dynamics determines the model's average Hamiltonian

$$\hat{H} = Tr[A], A = (\alpha_{it})_{i=1}^n \qquad (4.1)$$

where $\alpha_{it}$ is a local information speeds along a segment.

For these segments, being arranged in the IN's triplets, we get the average Hamiltonian

$$\hat{H} = \alpha_1(t_{1o}) \sum_{m=1}^{m=n/2} \gamma_2^{-m}, \qquad (4.1a)$$

which decreases with growing triplet's number, macromodel's dimension, and regular time (Sec. 3.2).

Each following triplet has less information speed than the previous one, and a total model's information speed is slowing down with growing a time of evolution.

Let us have a sum of the relative derivations of the model's eigenvalues spectrum at the current derivations:

$$D = \sum_{i=1}^n \frac{\Delta \alpha_i}{\alpha_i}, \qquad (4.2)$$

while the spectrum preserves *stability* at holding the VP.

Each relative *variation* of the spectrum in (4.2) is produced by the increment of correlations (Sec.3.1) according to the model's equations. We call (4.2) an *indicator* of model's *diversity*.

Using an average model's information speed, determined by an average Hamiltonian (4.1), we consider its relative increment:

$$H_\Delta = \frac{\Delta \hat{H}}{\hat{H}}, \Delta \hat{H} = \sum_{i=1}^{n} \Delta \alpha_{it} / \alpha_{it} \qquad (4.2a)$$

which, according to (4.2), is determined by the model diversity:

$$H_\Delta = D. \qquad (4.3)$$

Since $\alpha_{it} = \mathbf{a}(\gamma)\alpha_{io}$, $H_\Delta = H_\Delta(\Delta \alpha_{io}, \alpha_{io})$ and $H_\Delta$ is determined by initial diversity $D_o = \sum_{i=1}^{n} \Delta \alpha_{io} / \alpha_{io}$.

A maximal $\max D = D_o$ defines the spectrum's maximal variations, which are limited by a model's ability to preserve the spectrum dimension *n*. And $H_\Delta^o = D_o$ measures a maximal admissible increment of the average evolutionary speed ($H_\Delta^o$) for a *given* macrosystem, whose *n* dimensions we associates with a *population* of the macrosystem's interactive elements (subsystems).

A possibility of changing the initial relative $a_{io}^*(\gamma) = \Delta \alpha_{io} / \alpha_{io}$ under the stochastic inputs drives the model's variations. We evaluate a *maximum* among the admissible model's variations by a potential of evolution $P_e$:

$$P_e = \max \sum_{i=1}^{i=n} a_{io}^*(\gamma) = \max D_o \to \max H_\Delta^o. \qquad (4.4)$$

In the model's evolution, a limited $P_e$ counterbalances the admissible eigenvalue's deviations.

*Extending the population by increasing dimension n and growing the sum in (4.4) would raise the potential of evolution.*

The model limits the potential of evolution by its admissible maximal value $P_{em}$ for each m-th triplet, which preserves the triplet's dimension:

$$P_{em} = \max \alpha_{mo}^*(\gamma) \qquad (4.4a)$$

where

$$\alpha_{mo}^* = [(\alpha_{m2,o}^t)^2 - (\alpha_{m1,o}^t \alpha_{m3,o}^t)]^{1/2} / \alpha_{m3,o}^t = ((\alpha_{m2,o}^t / (\alpha_{m3,o}^t)^2 - \alpha_{m1,o}^t \alpha_{m3,o}^t / (\alpha_{m,o}^t)^2)^{1/2}$$
$$= (t_{m3}^2 / t_{m2}^2 - t_{m1} t_{m3} / t_{m3}^2)^{1/2} = [(\gamma_1^\alpha / \gamma_2^\alpha)^2 - (\gamma_2^\alpha)^{-1}]^{1/2} = \varepsilon(\gamma), \qquad (4.4b)$$

$$\gamma_1^\alpha(\gamma) = inv, \gamma_2^\alpha(\gamma) = inv, \alpha_{m1,o}^t t_{m1}^t = \alpha_{m2,o}^t t_{m2}^t = \alpha_{m3,o}^t t_{m3}^t = \mathbf{a}_o(\gamma) = inv,$$

and $\gamma_1^\alpha = t_{m2} / t_{m1}, \gamma_2^\alpha = t_{m3} / t_{m1}$ are the triplet eigenvalue's ratios, $t_{m1}^t, t_{m2}^t, t_{m3}^t$ are the segment's time intervals.

For a triplet, this ability will be compromised, if either $\alpha_{m2,o}^t$ approaches $\alpha_{m1,o}^t$, or $\alpha_{m2,o}^t$ approaches $\alpha_{m3,o}^t$,

*because* if both $|\alpha_{m2o}^t - \alpha_{im1,o}^t| \to 0, |\alpha_{m2o}^t - \alpha_{m3,o}^t| \to 0,$

the triplet disappears, and the macromodel's cooperative dimension *m* decreases.

Actually, the above minimal eigenvalue's distance is limited by the admissible minimal relative

$\alpha_{mo}^{*o}(\gamma^*) = \varepsilon_\gamma(\gamma^*)$ at minimal $\gamma^* = 0.007148$.

The optimal distance between both $\alpha_{m1,o}^t$ and $\alpha_{m2,o}^t$, $\alpha_{m2,o}^t$ and $\alpha_{m3,o}^t$ at the fixed $\alpha_{m1,o}^t$, $\alpha_{m3,o}^t$ satisfies to the known condition of dividing a segment with ($\alpha_{m1,o}^t, \alpha_{m2,o}^t, \alpha_{m3,o}^t$) in the mean and extreme ratio:

$$\frac{\alpha_{m1,o}^t}{\alpha_{m2o}^t} = \frac{\alpha_{m2o}^t}{\alpha_{m3,o}^t}, \qquad (4.4c)$$

which coincides with the above formulas for the triplet's invariant in (4.4b).

The triplet's eigenvalues ratio $\gamma_1^\alpha(\gamma^*) \cong 2.74, \gamma_2^\alpha(\gamma^*) \cong 4.9$, gives the maximal admissible error for total triplet $\varepsilon^2(\gamma^*) \cong 0.1086, \varepsilon(\gamma^*) \cong 0.33$.

The triplet's *information capacity* to counterbalance the maximal admissible deviations defines the triplet's potential $P_e^m$: $P_e^m = \max|\varepsilon(\gamma)|, \max\varepsilon(\gamma) = \varepsilon(\gamma^*)$. (4.5)

For an entire macromodel with $m$ such triplets, the total potential is

$$P_e^n = m P_e^m. \qquad (4.5a)$$

Thus, the macromodel's potential $P_e^n$ is limited by the maximal *acceptable* increment of dimension $n_m$ that sustains the macrostates' cooperation:

$$P_e^n \cong m_n/3, \ m_n = \frac{n_m - 1}{2}, \ P_e^n \cong (n_m - 1)/6n. \qquad (4.6)$$

This means, $P_e^n$ admits a maximal decrease of the model's current dimension $n_c$ in $n_m \cong n_c/3$ time.

Admissible interval of $\varepsilon(\gamma \in (0.001718 \to 0.8) \cong 0.33 \to 0.23$ determines an admissible interval of changing of the model potential in the ratio $P_e^* = (P_e^n(\gamma^*) - P_e^n(\gamma = 0.8)/P_e^n(\gamma^*)) \cong 0.3$. (4.6a)

A more realistic maximal potential $P_e^n(\min\gamma \cong 0.1) \cong 0.275,$ with the ratio

$P_{em}^* = (P_e^n(\min\gamma \cong 0.1)) - P_e^n(\gamma = 0.8))/P_e^n(\min\gamma \cong 0.1)) \cong 0.2$, decreases (4.6a) on 10% compared to $P_e^n(\gamma^*)$.

This leads to the admissible real change of the current dimensions: $n_{mr} \cong n_c/3.6$ that preserves the ratio (4.6) on a fewer diapason of $\gamma \in (0.1 \to 0.8)$. Within this diapason of $\gamma$, the average potential is $P_e^n(\gamma = 0.5)) \cong 0.256$.

Potential $P_e^n$ differs from $P_e$ (4.3), which generally does not support the evolution hierarchy of the process.

Relation $P_e^n \leq P_e$ limits the variations, acceptable by the model that restrict both the maximal increment of dimension and sustain the model's cooperative functions.

A *triplet's robustness* preserves the triplet's invariants under admissible maximal error $\varepsilon_m(\tau_{km})$, as a threshold for fluctuations, which satisfies the relation analogous to (4.4) at a current and fixed $\gamma = \gamma_*$:

$$P_{mr} = \max\varepsilon_m(\gamma_*), \varepsilon_m(\tau_{km}) = |\Delta\alpha_m(\tau_{km})/\alpha_m(\tau_{km})| \leq \varepsilon_m(\gamma_*), \qquad (4.7)$$

where $\tau_{km}$ is a window's time interval at forming $m$-th triplet,

External influences (mutations), affecting the error $\varepsilon_m(\tau_{km}) > \varepsilon_m(\gamma_*)$ within an admissible potential $P_{em}^*$, might change the model's invariants, dynamics and even the dimension, which could also altered the DSS code.

Let us show that at the equal deviations of the model parameter $\gamma_i^\alpha: \pm\Delta\gamma_1^\alpha, \pm\Delta\gamma_2^\alpha$, the model threshold $|\varepsilon(\Delta\gamma)|$ is asymmetric. The macromodel with $\gamma = 0.5$ has $\gamma_1^\alpha = 2.21, \gamma_2^\alpha = 1.76$ and get $\varepsilon_o = 0.255$.

Admissible deviations of $\Delta\gamma_1^\alpha = 0.25, \Delta\gamma_2^\alpha = 0.08$ correspond to the macromodel with $\gamma \cong 0.01$, $\gamma_1^\alpha = 0.246$, $\gamma_2^\alpha = 1.82$, which determines $\varepsilon_1 \cong 0.35$ and $\Delta\varepsilon(\Delta\gamma_1^\alpha, \Delta\gamma_2^\alpha) = \varepsilon_1 - \varepsilon_o = 0.095$.

At $\Delta\gamma_1^\alpha = -0.25$, $\Delta\gamma_2^\alpha = -0.08$ we have macromodel with $\gamma = 0.8$, $\gamma_1^\alpha = 1.96$, $\gamma_2^\alpha = 1.68$, which determines $\varepsilon_2 \cong 0.167$ and $\Delta\varepsilon(-\Delta\gamma_1^\alpha, -\Delta\gamma_2^\alpha) = \varepsilon_2 - \varepsilon_o = -0.088$.

It is seen that at the equal deviations, the model potential, defined by

$$\max \varepsilon = \max \varepsilon_1 \cong 0.35, \tag{4.7a}$$

tends to *increase at a decreasing* of $\gamma$, and vice versa.

An essential *asymmetry* of $|\varepsilon(\Delta\gamma)|$ and therefore the dissimilar $P_{em}$ are the results of the macromodel fundamental quality of *irreversibility*.

*The model's self-controllability and adaptivity.*

We assume that the VP's generated controls provides directional consolidation of the extremal segments' dynamics in a triplet's structures through, first, intervening each window between the segments and then consolidating the triplets in an IN.

Let's consider the model's optimal controls function

$$u_i(t_i) = \alpha_i^t v_i(t_i), \; v_i(t_i) = -2x_i(t_i), \tag{4.8}$$

applied at discrete moment $t_i$, where $x_i(t)$ satisfies to the controllable local differential equation for a segment (in a diagonal form): $\dot{x}_i(t) = -\alpha_i^t x_i(t)$. Then, the optimal control

$$u_k(t_i) = -2\alpha_k^t v_k(t_i) = -2\dot{x}_k(t_i), \tag{4.9}$$

can be applied directly using derivative $\dot{x}_k(t_i)$.

Considering a sequence of macrostates, satisfying the relations for the *phase* coordinates: $\dot{x}_i(t_i) = x_k(t_i)$ and $x_l(t_{i+1}), x_m(t_{i+2})$ in the subsequent discrete moments, we come to the macrostates' connections:

$$x_k(t_i) = \dot{x}_i(t_i), \; x_l(t_{i+1}) = \dot{x}_k(t_{i+1}), \; x_m(t_{i+2}) = \dot{x}_l(t_{i+2}). \tag{4.10}$$

If the macrostates' sequence has been *arranged* according to the model's time's course (Part 2), we come to relations

$$x_f(t_{i-1}) = x_i(t_{i-1}), x_k(t_i) = \dot{x}_i(t_i), x_m(t_{i+1}) = \ddot{x}_i(t_{i+1}), x_l(t_{i+2}) = \dot{x}_m(t_{i+2}) = \ddot{x}_i(t_{i+2}),...$$

which satisfy $x_{k-1}(t_i) = \dot{x}_k(t_i)$, or $x_{k-1}(\tau) = \dot{x}_k(\tau)$ for any following macrostate $k-1, k, k+1$ at fixed $t_i = \tau$.

This shows that each $x_{i-1}(t_i)$ is able to perform the *optimal control function*, when it's applied to each following $x_i(t_i)$ and generate the model optimal processes.

It requires using just initial macrostate $x_i(t_o)$ to be applied as a control $x_i(t_o) = u_{i+1}(t_o)$ to the equation

$$\dot{x}_{i+1}(t) = \alpha_{i+1,o}^t x_{i+1}(t) + u_{i+1}(t_o). \tag{4.11}$$

At $u_{i+1}(t_o) = \alpha_{i+1,o}^t x_{i+1}(t_o)$, we get the initial

$$\dot{x}_{i+1}(t_o + o) = 2\alpha_{i+1,o}^t x_{i+1}(t_o + o), \tag{4.11a}$$

which can be applied as a control

$$u_{i+1}(\tau_o) = -2\dot{x}_{i+1}(\tau_o) = -2\alpha_{i+1,o}^t x_{i+1}(\tau_o) \tag{4.11b}$$

at the moment $\tau_o$, following the initial $t_o$, where $t_o - \tau_o = \delta t_o$.

The sequence of the regular control's differences:

$$\delta u_{i+1}(\tau_o) = u_{i+1}(t_o) - u_{i+1}(\tau_o) = x_i(t_o) - 2\dot{x}_{i+1}(\tau_o), \tag{4.12}$$

applied at a small $\delta t_o$, forms the starting impulse control.

From the above relations follow the requirement

$$\alpha_{i+1,o}^t x_{i+1}(t_o) = \dot{x}_{i+1}(t_o) = x_i(t_o) = u_{i+1}(t_o). \tag{4.13}$$

If it holds, then optimal control $u_{i+1}(\tau_o) = -2\dot{x}_{i+1}(\tau_o)$ (4.11b), starting from the moment $\tau_o$, being applied to the right side of (4.13), will generate the model optimal processes sequentially at each moment $t_i, i = 1,..,n$.

Thus, arrangement of the model's initial conditions as the phase coordinates, satisfying relations (4.13) and (4.8), guaranties the model's self-controllability, which is performed by applying the starting optimal control.

Such natural processes that mutually control each other form a chain of *superimposing* processes (Lerner 1973).

In the considered model, implementation of self-controllability is possible *if* the state coordinates are connected by relation

$$x_{i-1}(t_o) = \alpha_{io}^t x_i(t_o) \tag{4.14}$$

at the initial moment $t_o$.

Then, as it has proven in (Lerner 1999), relation $x_{i-1}(t_i) = -2\dot{x}_i(t_i)$ would be held with the relative error

$$\varepsilon_i = |\delta x_i / x_i| = \exp(-\mathbf{a})\exp(\gamma_i^\alpha \mathbf{a}) - 2(1 - \mathbf{a}/\mathbf{a}_o) = inv(\gamma). \tag{4.15}$$

This error will satisfy the admissible in (4.7) if the model's self-control $u_i(t_i) = -2\alpha_i^t x_i(t_i)$, applied at each $t_i$, is able to achieve the self-controllability within an admissible threshold $\varepsilon_i \leq |\varepsilon_i(\Delta\gamma_i)|$ in (4.6), expressed via invariants $\mathbf{a}_o(\gamma), \mathbf{a}(\gamma)$ and the ratio of the nearest time intervals $\gamma_{i,i+1}^\alpha(\gamma) = \alpha_i / \alpha_{i+1} = t_{i+1}/t_i$.

With growing $\gamma$, the error (4.15) enlarges, and at $\gamma \to 1$, $|\frac{\mathbf{a}_o}{\mathbf{a}}| \to 1$, $\gamma_{i,m}^\alpha \to 1$ it reaches $\varepsilon \to 1$, when $\delta t_i \to t_i$, $t_{i+1} = t_i$ and the segment's dynamics repeat themselves. This breaks cooperation, which could continue at the non-coinciding eigenvalues holding the nearest $\gamma_{i,i+1}^\alpha > 1$; and in addition to that, the macromodel decays at $\gamma \to 1$.

Each of the optimal controls, applied at $t_i$, will be determined by the corresponding $(\alpha_{it} \pm \Delta\alpha_{it})$:

$$\Delta u_i^*(t_i) = \frac{\Delta u_i(t_i)}{u_i(t_i)} = \mp \frac{\Delta \alpha_i^t}{\alpha_i^t} = \alpha_{it}^*. \tag{4.16}$$

If $\alpha_{it}^*$ belongs to such *admissible* $\alpha_{i\tau}^*$, which preserves invariant $\mathbf{a}(\gamma)$, then this control also preserves invariants $\mathbf{a}_o(\gamma)$. These conditions keep $\gamma$ constant and, therefore, also preserves $\gamma_i^\alpha(\gamma)$ and hence provides $\mathbf{a}(\gamma)\gamma_i^\alpha(\gamma) = inv$ at the above admissible $\alpha_{i\tau}^*(\gamma)$, thereafter satisfying (4.15).

Such a self-control works *within* the model's *robustness* (4.7), which preservers $\alpha_{i\tau}^*$ at a fixed $\gamma = \gamma_*$.

Some extended self-controls enable *minimize* the maximal admissible triplet's error

$$\Delta t_i^*(\Delta\gamma) = 1 - \frac{t_i^2}{t_{i+1}t_{i-1}}(\gamma)$$ under the perturbations of $\alpha_{it}^*(\Delta\gamma)$ within *allowable* $\Delta\gamma \in \Delta\gamma_m$.

Such controls compensate the *acceptable* $\alpha_{it}^*(\Delta\gamma)$, *unrestricted* by the *robustness* potential $P_r$, but *limited* by $\max|\varepsilon_i(\Delta\gamma_i)|$ for any $\varepsilon_i \leq |\varepsilon_i(\Delta\gamma_i)|$ satisfying (4.15).

An optimal self-control, which automatically *adapts the model* to the changed $\alpha_{it}^*(\Delta\gamma)$ (within the admissible $\varepsilon_i \leq |\varepsilon_i(\Delta\gamma_i)|$) we call *adaptive* control.

This control's actions are limited by adaptive potential $P_a = \max|\varepsilon_i(\Delta\gamma_{io})|, \varepsilon_i \leq |\varepsilon_i(\Delta\gamma_i)|$, whose maximum is restricted by triplet potential $P_e^m$ : $P_a \leq P_e^m$, while $P_a$ provides both preservation of each triplet within the $n$ dimensional IN and adapts the acceptable variations not compromising the IN's hierarchy.

The adaptive controls support both *robustness* and adaptive potentials, which guarantees stability, preservation of the model triplets and total model's dimension $n$.

Essential *asymmetry* of $|\varepsilon(\Delta\gamma)|$ brings the related asymmetry to the adaptive potential.

At the variations within the $P_a$ capabilities, the generated mutations enable creating a new dimension with a trend to minimize both $\gamma$ and the system uncertainty (Sec.3.2).

The model's total adaptive potential $P_a^n = mP_a, m = n/2$ is an attribute of evolution dynamics, adapting allowable changes of the model's invariants and dimension. Extending the $P_a^n$ capability at growing $\gamma$ can bring instability, with a possibility to jeopardize the system cooperation.

The adaptive model's *function* can also be implemented by an external control's adaptive potential $P_a^n \leq P_e^n$.

The maximal increment of dimension, contributed to the adaptation, which also *restricts* the model's dimension.

### 4.5. *About possibility of model's reproduction*

The model's optimal control, which *preempts* and determines the process dynamics at each extremal segment, is memorized at the segment's starting time moment.

Such a memorized control *sequence* (as a code) can remember both the previous and the prognosis segment's dynamics. The model's DSS code, represented by the digits of these control sequence, is able to remember a preceding behavior of a complex system as a genetic code.

The DSS code, accumulated in the IN final node, memorizes the total process' dynamics and geometry and is able to reproduce them as the IN genetic code in new generated system.

An invariant information unit of this final node, measured by

$$h_{oo} = \mathbf{a}_o(\gamma) \tag{5.1}$$

defines an information *capability* needed to convey the IN genetics to a successor or other systems.

<u>Comments 4.5.</u>
A channel capacity in information theory $C$ limits information rate (in units of information per unit time) that can be achieved with an arbitrarily small error probability. The model's related channel capacity is defined by maximal information speed of its IN final node: $C = \alpha_m^N [bit/\sec]$.

For a given communication channel with fixed channel capacity $C_o$, transmission through this channel a total information, measured by the integral functional $\Delta S[\tilde{x}(t)]$, requires the time interval $T_o = \Delta S[\tilde{x}(t)]/C_o$, which is longer than that for transmission of the same process' $\tilde{x}(t)$ information, encoded by traditional Shannon's information measure $\Delta S_o$, since $\Delta S[\tilde{x}(t)] > \Delta S_o$ includes hidden information of the same process, Parts 1, 2. •

Decoding the IN genetic code allows restoration of the specific position of each node within the IN structure, the process' invariants and the optimal control. Potential newborn system sequentially evolves as a new IN structure that encloses the system's dynamics, geometry and the related DSS code.

A behavioral set of the IN genetic code, which had memorized the previous behavior, could reproduce it in the successor. That is why a successor, possessing this code, is able to *anticipate and preempt* its behavior including creation of the reproductive dynamics with *more valuable information.*

A successor's process of reproduction could start by acquiring just an initial triplet, which is able to generate its IN. Such a triplet requires three starting digits from potentially two "parents", completed of two digits from the IN final node of one parent and one starting digit from another parent's IN final node.

Each of these IN's final nodes could carry instability, since within each IN, an internal information, transferred to a following triplet's node, is compensated, except the IN final node.

At forming a daughter's triplet, such instability would be compensated by assembling all obtained three digits in its starting IN's node. The parents' triplet ought to be prepared through consolidating the two encoded eigenvalues of one parent with a third one from other parent. This requires, first, enfolding these two unstable eigenvalues into a single stable unit, which then will be equalized with the unstable third one that must have the same equal eigenvalue at the consolidation. Thus, for a successful process of reproduction, the parent, who delivers two encoded eigenvalues, should prepare them for both being stable and equal with the eigenvalue, delivered by other parent.

Other parent of such reproductive process needs to carry the encoded eigenvalue being equal to that delivered with two eigenvalues above. Such requirements to both parents limit the successful reproduction.

Three digits of the parents code starts forming a "daughter's" IN node, reproducing the genetic information, had received from the parents. Depending on information values of each parent's code digits, this node could reproduce any of the parent's genetics: either from just one of them, or according to their mixed code's sequence.

In a complex system, an *association* of INs with different $(n, \gamma)$ can mutually interact with the negentropy-entropy exchanges between them analogous to a crossover. The ranged IN's frequencies could be wrapped up in each other changing the IN's invariants and inner codes. Such a system might create the association's common code, as a genetic coding sequence, received from the *final* nodes of various INs, and then be a source of its "parent's code".

Such a parent's code with its particular variations is a *common source* of potential evolutionary changes on the IN's higher (intelligence) level.

The considered below a cyclic process of a reproduction (Sec.3.5), involves transferring an eigenvalue of the IN's last node to an eigenvalue of new starting dynamic process, which is able to produce new triplet and then the related IN. It has shown that in such transformation, the ratios of nearest two eigenvalues $\alpha_{n-1}/\alpha_{1o} = \gamma_1^\alpha, \alpha_{n-1}/\alpha_{3o} = \gamma_2^\alpha$ keep their values satisfying the VP.

According to the Second Law, each following cycles has a tendency of fading and chaotization in periodical processes. In addition, *random* initial eigenvalues, generated at the beginning of the cycle, *do not replicate* the ones at the cycle end.

Formal analysis of the cycle renovation (Sec.3.6) is associated with a nonlinear fluctuation, which can be represented (Kolmogorov1978) by a superposition of linear fluctuations with the frequency spectrum ($\omega_1^*,...,\omega_m^*$) proportional to the imaginary components of eigenvalues ($\beta_1^*,...,\beta_m^*$), where $\omega_1^*$ and $\omega_m^*$ are the minimal and maximal frequencies of the spectrum accordingly. Such nonlinear fluctuations are able to generate a new model with new real and imaginary eigenvalues, which can give a start to a newborn macroprocess and the following IN cooperative macrodynamics. This leads to the model's *cyclic functioning*, initiated by two mutual controllable processes, which should not consolidate by the moment of the cycle renovation.

In such a cyclic macromodel functioning, after the model disintegration, the process can repeat itself with the state integration and the transformation of the imaginary into the real information during the dissipative fluctuations.

The model of the cyclic functioning includes the generator of random parameters that renovates the macromodel characteristics and peculiarities, constrained by the VP (Lerner 2001). The adaptive potential in this cycle increases under the actions of the adaptive control even if the randomness carries a positive entropy. Such an adaptive, repeating, self-organizing process is the *evolutionary cycle* with potentially growing valuable information.

In the evolutionary cycle, a code, possessing the information channel capability (5.1), might transfer the DSS genetics to others using for communication a wave frequency of the model's final eigenfunction (Part 1).

The above conditions for existence of the adaptation potential and the model's cyclic renovation impose a *limitation* on the minimal admissible macromodel's *diversity*. With similar reproductive dynamics for all population, the reproduction, applied to a set of the macromodels, having a maximum diversity, brings a maximum of adaptive potential, and therefore, it would be more adaptive and beneficial for all set, potentially spreading through the population. Such an evolution trend is confirmed in a latest publication (Greig 2008).

The structural stability, imposed by the VP and implemented in the IN, affects restoration of the system structure in the cyclic process through the process reproduction.

The macrosystem, which is able to continue its life process by renewing the cycle, has to transfer its coding life program into the new generated macrosystems and provide their secured mutual functioning.

### 3.6. *Mathematical specifics of cyclic evolution*

*Proposition* 6.1. A nonlinear fluctuation is able to generate a new model with the parameter

$$\gamma_{lo} = \frac{\beta_{lo}(t_o)}{\alpha_{lo}(t_o)} = \frac{\beta_n^i(t_{n+k})}{\alpha_n^i(t_{n+k})} = \frac{2\cos(\beta_{n+k-1}^i t) - 1}{2\sin(\beta_{n+k-1}^i t)}, \tag{6.1}$$

where $\alpha_{lo}(t_o) = \alpha_i^*(t_{n+k})$, $\beta_{lo}(t_o) = \beta_i^*(t_{n+k})$ are the new model's starting real and imaginary eigenvalues.

*Proof.* During the oscillations, initiated by the control action, a component $\beta_i^*$ is selected from the model's imaginary eigenvalues $\text{Im}\lambda_n^i(t) = \text{Im}[-\lambda_{n-1}^i(2-\exp\lambda_n^i t)^{-1}]$, at each $t = (t_{n+k-1}, t_{n+k})$.

We come to relation

$$\operatorname{Im} \lambda_n^i(t_{n+k}) = j\beta_n^i(t_{n+k}) = -j\beta_{n+k-1}^i \frac{\cos(\beta_{n+k-1}^i t) - j\sin(\beta_{n+k-1}^i t)}{2 - \cos(\beta_{n+k-1}^i t) + j\sin(\beta_{n+k-1}^i t)}, \quad (6.2)$$

at $\beta_i^* = \beta_{n+k}^i, \beta_i^* \neq 0 \pm \pi k$. It seen that $\beta_n^i(t_{n+k})$ includes a real component

$$\alpha_n^i(t_{n+k}) = -\beta_{n+k-1}^i \frac{2\sin(\beta_{n+k-1}^i t)}{(2 - \cos(\beta_{n+k-1}^i t))^2 + \sin^2(\beta_{n+k-1}^i t)}, \quad (6.3)$$

at $\alpha_i^* = \alpha_i^*(t_{n+k}) \neq 0$, with the corresponding parameter

$$\gamma_i^* = \frac{\beta_n^i(t_{n+k})}{\alpha_n^i(t_{n+k})} = \frac{2\cos(\beta_{n+k-1}^i t) - 1}{2\sin(\beta_{n+k-1}^i t)}. \quad (6.4)$$

These eigenvalues $\lambda_i^*(t_{n+k}) = \alpha_i^*(t_{n+k}) \pm j\beta_i^*(t_{n+k})$, at some moment $t_o > t_n$, could give a start to a new forming macromodel with $\lambda_{lo} = \alpha_{lo}(t_o) \pm j\beta_{lo}(t_o)$, with the initial real $\alpha_{lo}(t_o) = \alpha_i^*(t_{n+k})$, the imaginary $\beta_{lo}(t_o) = \beta_i^*(t_{n+k})$, and the parameter $\gamma_{lo} = \frac{\beta_{lo}(t_o)}{\alpha_{lo}(t_o)}$ equals to (6.1). •

*Comments* 6.1. This new born macromodel might continue the consolidation process of its eigenvalues. Therefore, returning to some primary model's eigenvalues and repeating the cooperative process is a quite possible after ending the preceding consolidations and arising the periodical movements. This leads to the cyclic macromodel functioning when the state integration alternates with the state disintegration and the system decay. (The time of the macrosystem decay increases with an increase of the accuracy $\varepsilon^* = \frac{\Delta x_n}{x_n}$ of reaching the given final state).

Since the system instability occurs at $\gamma \geq 1$ (while at $\gamma \to 0$ the cooperative process begins), we might associate the cycle *start* after its ending on the invariant loop (Fig.3.1) at $\gamma \geq 1$ with the model's oscillations (generating the spectrum $\lambda_i^*(t_{n+k})$). Then the end of such cycle is following the start of the newborn macromodel at $\gamma \to 0$.

In this case, $\gamma \to 0$ can be achieved in (6.4) at $2\cos(\beta_{n+k-1}^i t) \to 1$, or at
$(\beta_{n+k-1}^i t) \to (\pi/3 \pm \pi k), k = 1,2,...$, with $\alpha_n^i(t_{n+k}) = \alpha_{lo}^m(t_o) = \lambda_{lo}^m \cong -0.577\beta_{n+k-1}^i, \beta_{lo}(t_o) \cong 0$,
whereas $\gamma = 1$ corresponds to $(\beta_{n+k-1}^i t) \approx 0.423 rad (24.267^o)$, with $\beta_i^*(t_{n+k}) \cong -0.6\beta_{n+k-1}^i$.

Here $\beta_i^*(t_{n+k}) \cong \alpha_{lo}^m(t_o)$ determines the maximal frequency $\omega_m^*$ of the fluctuation by the end of the optimal movement. And the new macromovement starts with this initial frequency.

*Proposition* 6.2. The maximal *frequency's ratio*, generated by the initial (n-1) dimensional spectrum with an imaginary eigenvalue $\beta_{n-1,o}(t_{n-1,o})$ by the end of the cooperative movement at ($\gamma = 1$):

$$\frac{\beta_i^*(t_{n+k})}{\beta_{n-1,o}(t_{n-1,o})} = l_{n-1}^m \quad (6.5)$$

is *estimated by the invariant*

$$\frac{0.577\pi/3}{a_o(\gamma)/a(\gamma) \ln 2} = l_{n-1}^m (\gamma = 1). \quad (6.6)$$

*Proof.* Let us estimate (6.5) using the following relations.

Because $\beta_{n-1,o}(t_{n-1,o}) = \alpha_{n-1,o}(t_{n-1,o})$ at $\gamma = 1$ and $\beta_i^*(t_{n+k}) \cong \alpha_{lo}^m(t_o)$, we come to $l_{n-1}^m = \frac{\alpha_{lo}^m(t_o)}{\alpha_{n-1,o}(t_{n-1,o})}$.

Applying relations $\alpha_{n-1,o}(t_{n-1,o}) = \alpha_{n-1,t}(t_{n-1}) \mathbf{a}_o(\gamma)/\mathbf{a}(\gamma)$ and using $\alpha_{lo}^m(t_o) = -0.577\pi/3$, we have

$$\frac{\alpha_{lo}^m(t_o)t_o}{\alpha_{n-1,o}(t_{n-1})t_o} = \frac{0.577\pi/3}{\mathbf{a}_o(\gamma)/\mathbf{a}(\gamma)\alpha_{n-1,t}(t_{n-1})t_o} = l_{n-1}^m(\gamma = 1). \qquad (6.7)$$

Assuming that a minimal $t_o$ starts at $t_o = t_{n-1} + o(t) \cong t_{n-1}$, and using invariant $\mathbf{a}_o(\gamma = 0.5) = \alpha_{n-1,t}(t_{n-1})t_o = \ln 2$, at $\alpha_{n-1,t}(t_{n-1}) = \alpha_{n-1}$, we get the invariant relation (6.6). •

<u>Comments</u> 6.2. The ratio (6.6) at $\mathbf{a}_o(\gamma=1)=0.58767$, $\mathbf{a}(\gamma=1)=0.29$ brings $l_{n-1}^m(\gamma=1) \approx 0.42976$.

Because $\alpha_{n-1,o}(t_{n-1,o})$ is the initial eigenvalue, generating a starting eigenvalue $\alpha_{lo}^m(t_o)$ of a new model, their ratio $\frac{\alpha_{n-1,o}(t_{n-1,o})}{\alpha_{lo}^m(t_o)} = \gamma_l^{\alpha m}$ determines the parameter of the eigenvalue's multiplication for the new formed model, equal to $\gamma_l^{\alpha m} = (l_{n-1}^m)^{-1}$, and at $\gamma = 1$, we get $\gamma_l^{\alpha m} \approx 2.327$. For the new model, the parameter $\gamma_l \to 0$, at which the cooperative process and the following evolutionary development might start, holds true. For a new formed triplet, satisfying relations (4.4b,c), we get the eigenvalues and their parameters of multiplication in the forms:

$\alpha_{l-1,o}^m(t_o) = \alpha_{n-1,o}(t_{n-1,o}) \approx -1.406$, $\alpha_{lo}^m(t_o) \approx -0.60423$, $\alpha_{l+1,o}^m(t_o) \approx -0.26$, with

$$\gamma_{l-1}^{\alpha m}(\gamma) = \frac{\alpha_{l,o}^m(t_o)}{\alpha_{l+1,o}(t_o)} \approx 2.3296 \text{ and } \gamma_{l-1,l+1}^{\alpha m}(\gamma) = \frac{\alpha_{l-1,o}^m(t_o)}{\alpha_{l+1,o}(t_o)} \approx 5.423. \qquad (6.8)$$

If the optimal cooperation in the new formed model is continued, both parameters $\gamma_l$ and $\gamma_l^{\alpha m}$ will be preserved, which determines the new model's invariants and information genetic code.

The transferred invariants are the carriers of the evolutionary hierarchical organization, self-control, and adaptation (while the adaptation process could change both $\gamma_l$ and $\gamma_l^{\alpha m}$). •

### *3.7. The main evolutionary regularities and a unified law of evolution*

We finalize above results, including Parts1-2, by following formulation the regularities and unified information law.

1.*The mathematical law of creation the dynamic regularities from stochastics* is defined by the dependency of the model's both kinetic operator $L_k$ and a local *gradient of dynamic potential* $gradX(\tau)$ on a dispersion matrix $b$ in the form

$$L_k(x, \tau+o) = 1/2 b(x, \tau+o), \quad (\int_{\tau-o}^{\tau} b(t)dt = 2M[\tilde{x}(\tau)\tilde{x}(\tau)^T]] \qquad (7.1)$$

$$gradX(\tau+o) = (\int_{\tau-0}^{\tau+o} \sigma\sigma^T dt)^{-1}, \sigma\sigma^T = 1/2b, \qquad (7.2)$$

which defines also the related drift vector in the entropy functional (Part1):

$$a^u(\tau+o) = \sigma\sigma^T(\tau+o)X(\tau+o) \qquad (7.3)$$

and the model's dynamic operator $A$, while the gradient of dynamic potential (7.2) connects the dynamics and stochastics. Since the diffusion part in (7.2) tends to decrease (Sec.3.1), the gradient (7.2) grows, which in turn diminishes an evolution impact of randomness on the dynamics, and vice versa.

The dynamic potential, which exists only along the extremal segments, depends on the diffusion, identified at the punched localities between the segments.

The Eqs (7.1-7.3) connect the dynamics and stochastics, revealing the discrete localities (DP) where the dynamics inherit the randomness and then retain them in the information invariants; the invariants determine the dynamic movement, renovating at each following extremal segments at these localities.

2. *The law for the gradient of dynamic potential, local evolutionary speeds, the evolutionary conditions of a fitness, and a lifetime*.

At the beginning of each extremal, the gradient of potential reaches a local maximum (as the impulse control dissolves correlations in (7.1)), and then evolves on the extremal in the inverse proportion to the correlation *on* the extremal (following Sec.3.1):

$$gradX(t) = (E[x(t)x(t)^T])^{-1}. \qquad (7.4)$$

Considering each extremal movement $x(t)$ within the segment time interval $t_k = (t_k^i)$ (between the moment $t_o \leq t < t_k$) and using the equation for correlation function $r(t)$ on an extremal:

$$E[x(t)x(t)^T] = E[x(t_o)x(t_o)^T]exp(2At), A = A^T, A < 0, t_o = \tau + o, \tau = \{\tau_k\}, t_o = \{t_o^i\}, t_k = \{t_k^i\}, i = 1,...,n, n \geq m,$$

we get the gradient (7.4) by the moment $t_k$ in the form

$$gradX(t_k) = gradX(t_o)exp(2At_k), gradX(t_o) = gradX(\tau + o). \qquad (7.5)$$

This means that, at a stable extremal movement (with $A < 0$), the local maximum of the gradient (at the DP, corresponding $A(\tau) > 0$) declines, reaching the local minimums by the segment's end according to (7.5).

Along the evolution dynamics, represented by a chain of the extremals, the local minimums of the potential's gradients are diminishing, as the related local eigenvalues are lessening (according to Prop.2.2.1, Part 2).

During the extremal movement, the entropy derivation $dS/dt$ (which *directs* the evolution dynamics), starting with a complex conjugated eigenvalues $|\lambda_i(t)|$ is reduced, approaching the moment of imposing the dynamic constraint with a real eigenvalues (at $t = t_k$, $\beta_i(t_k) = 0$). The real eigenvalues $\alpha_i(t)$, satisfying a minima of the derivation at any $t < t_k, \alpha_{io} < 0$, decline, approaching a minimum at $t = t_k$:

$$\min E \mid \frac{\partial S_{ik}}{\partial t} \mid \to \min \mid \alpha_i(t_k^i) \mid, i = 1,...,n. \qquad (7.6)$$

These minimus at the end of each extremal segment in the form

$$\min \alpha_i(t_k^i) = \min \alpha_j(t_k^j), (\alpha_i(t_k^i), \alpha_j(t_k^j)) = diagA(t_k), \qquad (7.7)$$

allows joining the extremals in the IN (Fig.2.4) with the local information speeds, satisfying (7.6,7.7).

Such a cooperation is constrained by the requirement of connecting the eigenvalues (eigenvectors) of the *same* matrix *A* according to relations (4b,c) for a triplet.

From (7.4, 7.5), it follows that at negative *A*, the gradient is negative on the extremal.

Relations (7.5)-(7.7) lead to the invariant condition for the gradient at the cooperating segments' ending moments:

$$\min \mid \frac{d}{dt}[gradX_i(t_k^{i-1})]\{[gradX(t_k^{i-1})]\}^{-1} \mid = \min \mid \frac{d}{dt}[gradX_i(t_k^i)]\{[gradX(t_k^i)]\}^{-1}, i = 1,...n. \quad (7.8)$$

These conditions, connecting the *local* process' information *speeds* to satisfy the above specific triplet's ratios, express the requirement for the VP extremals to *fit e*ach other in the *evolutionary cooperative* dynamics.

While each starting local maxima of the dynamic gradient intensifies an impact of the following dynamics on the evolutionary development, its decreasing value by the end of each segment, diminishes such influence.

The law for an *average information speed* along the trajectory of *evolutionary dynamics* (7.6) in the form:

$$E \mid \frac{\partial S(t_k)}{\partial t} \mid = \mid \sum_{i=1,k=1}^{n,m} \alpha_i(t_k^i) \mid \rightarrow \min, \qquad (7.9)$$

depends on the entire spectrum's eigenvalues, or a total dynamic potential, representing the whole system (as a population), which accumulates an emergence of the random features for a whole population.

The speed grows with adding each following eigenvalue in cooperative process, even though this eigenvalue decreases, while the local speed (7.6), measured by a *decreased eigenvalue* in the matrix trace (7.9) could weaken each local evolutionary information dynamics.

A minimal *acceptable* eigenvalue for the cooperation, corresponding to a local maximum of the dynamic gradient, limits the evolutionary speed. The evolution speed grows with enlarging dimension of a population.

The evolutionary model possesses two scales of time (Sec.3.2). Both these times evolve with evolution of the invariants and the system information dynamics. However, the optimal process evolves according to the *regular* time course with growing dimensions of its cooperative structures and their lifetime.

4. *The law for diversity, variations, stability, a potential of evolution, and complexity*.

The relative increment of the averaged information speed (7.9):

$$E \mid \frac{\partial S_i^*}{\partial t} \mid = 4 \sum_{i=1}^{i=n_i \pm \Delta n_i} \mid \frac{\Delta \alpha_{io}}{\alpha_{io}} \mid, \alpha_{io} = \operatorname{Re} \lambda_{io}; D = \sum_{i=1}^{i=n_i \pm \Delta n_i} a_{io}^*(n,\gamma), a_{io}^* = \mid \frac{\Delta \alpha_{io}}{\alpha_{io}} \mid. \qquad (7.10)$$

could be measured by diversity $D$ (4.2) as the sum of the *admissible variations* for the system's eigenvalues spectrum, which *preserves* the spectrum *stability* at current derivations. Whereas each variation is produced by the increment of diffusion according to (7.1), or the related dynamic potential (7.2).

A max $D = D_o$ defines the spectrum's maximal variation, limited by a system's ability to sustain the spectrum extended *dimension* $n_{im} = n_i \pm \Delta n_i$. Relation $P_e = \max D_o \rightarrow \max H_\Delta^o \qquad (7.11)$ measures a maximal potential of evolution $P_e$ and brings a maximal increment of the average information evolutionary speed $H_\Delta^o$:

$$H_\Delta^o = \max E \mid \frac{\partial S_i^*}{\partial t} \mid \qquad (7.12)$$

for a given macrosystem's dimension, related to the size of a population. Extending the population's size by increasing its dimension $n$ (satisfying the stability) raises the potential of evolution.

Potential $P_e$ (7.11), (Sec 3.4) generally does not support the evolution hierarchy of the process.

The model potential $P_e^n$ (4.3) that sustains the macromodel's cooperation admits decreasing its current dimension $n_c$ in ratio $n_{mr} \cong n_c / 3.6$. Relation $P_e^n \leq P_e$ limits the variations, acceptable by the model that restricts both the maximal increment of dimension and sustains the model's cooperative functions.

The evolution dynamics, accompanied by arranging the extremal segments in an ordered sequence, leads to sequential formation of cooperative structures, assembling in the cooperative's hierarchical organization.

The cooperative information complexity, defined by a ratio of the average information speed (4.12) in the cooperative evolution dynamics *to* a cooperated volume (Sec.2.7, Part 2), is a *collective information measure* of evolving cooperative, which depends on the numbers of information elements (structures) assembling in the cooperative. At assembling new information elements in a cooperative space distributed hierarchical system (IN), each IN's current structure wraps and absorbs the complexity of all previously joint IN's structure.

Because of that, the cooperative complexity de*creases* with growing the dimension of a cooperating system.

The speed of each following evolutionary cooperation, related to a previous formation, also *decreases*.

Above evolution regularities emerge at the model's DP punched localities, where each local variance is fixed by the cooperative binding and it is encoded in the DSS.

5. *The law of evolutionary hierarchy, potential adaptation, adaptive self-controls and a self-organization; coping, genetic code, and the error correction.*

The evolutionary dynamics, created by the multidimensional eigenvalues' spectrum, forms a chain of interacting extremal segments, which are assembled in an *ordered space-time organization* structure of the information hierarchical network (IN).

The requirements of preserving the evolutionary hierarchy (4.4b,c) impose the restrictions on the maximal potential of evolution $P_e$ and limit the variations, *acceptable* by the model.

*The evolution process enables self-controllability and adaptivity* (Sec.3.4) .

The adaptive model's *function* is implemented by the adaptive *feedback-control*, acting within the $P_a$ capabilities.

The asymmetrical adaptive potential leads to both rising the system's adaptive capability (at decreasing $\gamma$) with expanding the potential ability to correct a current error *and* to increasing of the impact of a dynamic prehistory on current changes. This asymmetry contributes the model's evolutionary improvement.

The self-control function is determined by the conditions (4.12-4.16) of a proper *coordination* for a chain of superimposing processes, where each preceding process adopts the acceptable variations of each following process.

The above optimal controls are synthesized, as an *inner feedback*, by the *duplication* of and *copying* the current macrostates at the beginning of each segment, which are *memorized* and *applied* to the segment.

The space distributed IN's *structural robustness* is preserved by the feedback actions of the inner controls, which provide *a local stability at the admissible* variations.

This control supports a limited $P_e(\gamma) = P_r$ that determines the *potential of robustness.*

An individual macromodel *lifetime* (Sec. 3.2 and Part 2) is *limited* by both the number *n* of the ranged eigenvalues (the local entropy's speeds) and a maximal value of these entropies.

At growing these information speeds and preserving the system's invariants, the macromodel's *lifetime is shortening*, depending only on the entropy speed of the IN's final cooperating segment.

A sequence of the sequentially enclosed IN's *nodes*, represented by a *discrete control logic*, creates the IN's *code* as a *virtual communication language and* an algorithm of minimal program to *design* the IN.

The optimal IN's code has a *double spiral* triplet structure (DSS), shaped at the localities of the sequential connected cones' spirals (Fig.2.5d), which form the time-space path-line of transferring the IN's information through the triplet's hierarchy. The applied control adds the forth letter to the initial minimal three triplet's code letters, which provides the model's *error correction mechanism* to the IN and its DSS code; it also provides discrete filtering of the randomness, acting at the triplet's window (Part 2)**.**

The IN information geometry holds the node's binding functions and an asymmetry of triplet's structures.

In the DSS information geometry, these binding functions are encoded, in addition to the encoded nodes' dynamic information. The DSS specialization depends on the structure of the random process' functions drift and diffusion (Parts 1,2). The IN's geometrical border forms an external surface Fig.2.6, where the macromodel is open for the outside interactions, and the interacting states and nodes *compete* for delivering a maximum of information.

Selected states are copied and memorized by the model's control, contributing to the code.

The IN's evolutionary dynamics accumulate the cooperative contributions, leading to a renovation of the information invariants and DSS code. The control provides a *directional* evolution with the extraction of maximum information from the environment of competing systems, while the acquired DSS code can be passed to a successor (Sec.3.5). The IMD software packet (Lerner 2010) simulates the main mechanisms of the evolutionary dynamics.

**Conclusion**

The regularities of evolution dynamics, following from VP, hold the following main *specifics*:

Tendency of *diminishing* the influence of randomness on the macrolevel dynamics;

Decrease the contributions from diffusion with the increase of the dynamic gradient that intensifies a *growing impact of the each following dynamics* on the evolutionary development;

An *attracting* cooperative impact from particular cooperative dynamics that drives evolutionary cooperation;

Decline of an average cooperative evolutionary information speed (for a population of subsystems) during the time of evolution with a limited individual lifetime;

Nonsymmetrical adaptive potential that leads to both raising the system's adaptive capability with expanding the potential ability to correct a current error *and* increasing of the impact of a dynamic prehistory on current changes;

Forming the cooperative organization structures, ordered in the evolutionary hierarchy of information dynamic networks with growing the values and density of collected information and decreasing the cooperative complexity along the hierarchy. Such organization structures are capable of adaptive self-functioning, which limits their *minimal* nodes number (units) by eight.

More complex organizations are composed from a set of the minimal units, collected from single and/or different discrete ensembles, which satisfy the conditions of their cooperation.

Therefore, *emergence of such organizations is possible by discrete evolutionary steps, having an irreproducible hierarchical cooperative complexity*.

The limitations increase the *information differences* between complexities of the nearest IN levels' subsystems, *restricting* formation such of them, which unable support the level's cooperative dynamics.

The evolutionary changes in the growing structural organizations, which support the cooperation, are declining, and on the IN's highest level are possible only through *particular variations of its genetic DSS code.*

The VP is *a single form of mathematical law that defines the above regularities, which prognosis the evolutionary dynamics and its specific components: potential, diversity, speed, complexity, and evolving genetic code.*

Such a generalized law that embraces all above regularities consists of *maximizing a minimum* of available information along the evolution process, which establishes both optimal collection of current information and its optimal evolution dynamics with above regularities.

By observing a random environment with a variety of natural processes, an *observer* extracts random information and converts it to its internal dynamic processes.

Generally, an information observer is a *formal performer of the VP* through emerging interactions in the observing process, which produce a sharp impact, analogous to the impulse control's actions.

Each interaction sequentially transforms the EF random information in the IPF partitioned dynamic (including quantum) information. It could also be a result of a quantum information entanglement at the widow locality, generated by a decay of correlation functions from a sharp natural or artificial impact.

Information of connected–s*uperimposed* states of a process is quantum information of an entanglement.

The entanglement produces a *quant* (qubit) of the entangled *dynamic* states' information, which sets up an elementary *observer* and generates both local and nonlocal attractions, enable adjoin new information.

The *elementary* information observer, as a possessor of qubit from the superposition of two-state quantum system in a *multi-state* system, could emerge from any previous information extraction via an impulse controls action (or interactions) (as it has shown in Lerner 2012a). Thus, the *observed information creates the observer.*

By attracting new quantum information, the observer starts elementary cooperatives of its inner information structure. The observer's cooperative information dynamics, developing under the law from diffusion stochastics, build the evolving cooperative structures with a sequentially growing their hierarchical organizations, whose genetic code enables a cyclic reproduction and progressive evolution. The cooperative information geometry, evolving under observations, limits the size and volumes of a particular observer. Hence, *information drives evolution*.

The IN's highest level of the structural hierarchy evaluates the observer's *intelligence level*, whose evolution is determined by the VP *dynamic* law rather than random environment, its impact during the evolution *diminishes*.

In the diffusion process, such as random trajectory of the diffusing particles, the hidden information is materialized by the particles' elementary interactions. And the law's evolution cooperative dynamics will be implemented through real physical, chemical, biological structures, which for highly organized systems would create a *cognition and intelligence of an information observer* (Lerner 2012b).

**References to Part3**

**Figures from Sec.2 which are referred in Part3:**

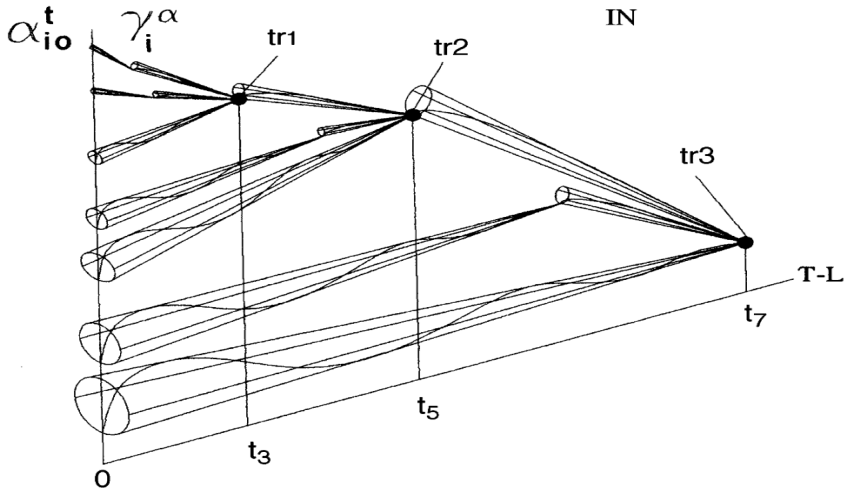

**Fig. 2. 4. The IN time-space information structure, represented by the hierarchy of the IN cones' spiral space-time dynamics with the triplet node's (tr1, tr2, tr3, ..), formed at the localities of the triple cones vertexes' intersections, where $\{\alpha_{io}^t\}$ is a ranged string of the initial eigenvalues, cooperating around the $(t_1, t_2, t_3)$ locations; T-L is a time-space.**

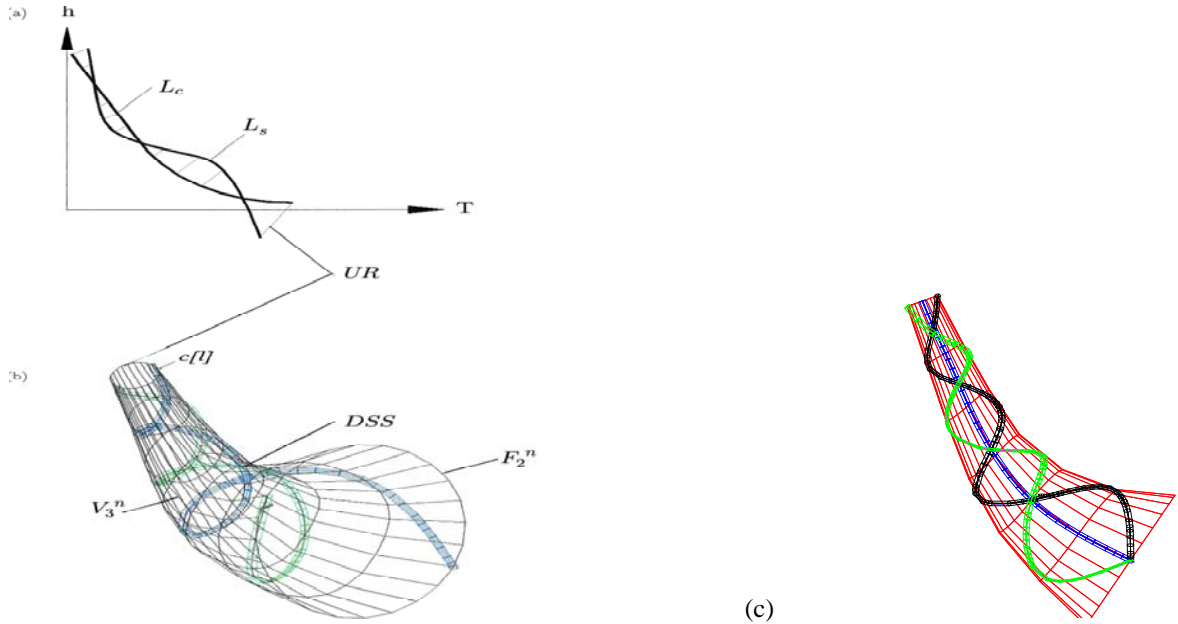

(c)

**Fig 2.5. (a) Simulation of the double spiral cone's structure (*DSS*) with the cell (c[l]), arising along the switching control line *Lc*; with a surface $F_2^n$ of uncertainty zone (*UR*) (b), surrounding the *Lc*-hyperbola in the form of the *Ls*-line, which in the space geometry enfolds a volume $V_3^n$ (b,c); (c) Simulation of the double spiral code's structure (*DSS*), generated by the IN's nodes: the central line models the IN node's information cells; the left and right spirals encode the IN's states, chosen at the DPs by the IC control's double actions.**

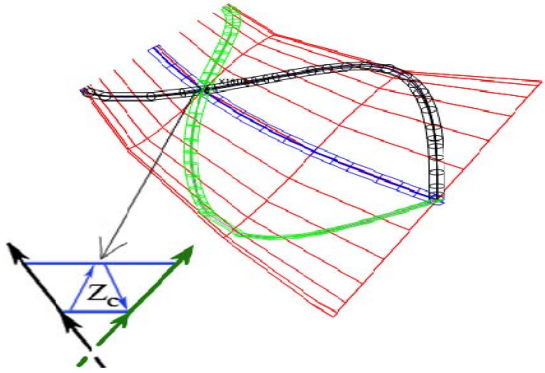

**Fig.2.5d. Zone of cells $Z_c$, formed on the intersections of opposite directional spirals, which produces each triplet's DSS code.**

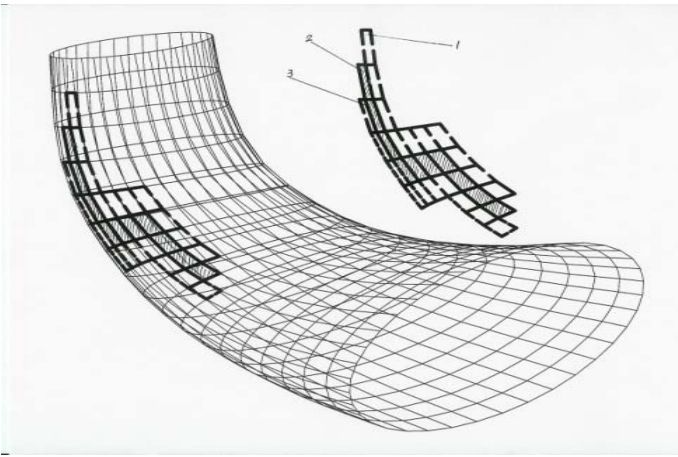

**Fig. 2. 6. Structure of the cellular geometry, formed by the cells of the DSS triplet's code, with a portion of the surface cells (1-2-3), illustrating the space formation.**